\documentclass{emulateapj}

\usepackage{amsmath}
\usepackage{cases}
\usepackage{txfonts}
\usepackage{tipa}
\usepackage{amsfonts}
\usepackage{amssymb}
\usepackage{graphicx}
\usepackage{multirow}
\usepackage{bigstrut}
\usepackage{array}

\shortauthors{Gu et al.}

\begin{document}

\title{Galactic Stellar Populations from Photometric Metallicity Distribution Functions}

\author{Jiayin Gu\altaffilmark{1}, Cuihua Du\altaffilmark{2},  and  Wenbo Zuo\altaffilmark{3}}
\affil{$^{1}$ Department of Physics, Wuhan University of Technology, Wuhan 430000, P. R. China; gujiayin12@mails.ucas.ac.cn \\ 
$^{2}$ College of Astronomy and Space Sciences, University of Chinese Academy of Sciences, Beijing 100049, P. R. China; ducuihua@ucas.ac.cn \\
$^{3}$ Key Laboratory of Optical Astronomy, National Astronomical Observatories, Chinese Academy of Sciences, Beijing 100101, P. R. China}

\begin{abstract}
Based on Sloan Digital Sky Survey (SDSS) photometric data, Gu developed a new Monte-Carlo-based method for estimating the stellar metallicity distribution functions (MDFs). This method enables a more reliable determination of MDFs compared with the conventional polynomial-based methods. In this work,  MDF determined from the method are well fit by three-Gaussian model, with peaks at ${\rm [Fe/H]}$=$-0.68$, $-1.38$, and $-1.90$, associated with the thick disk, inner halo, and outer halo, respectively. The vertical metallicity gradient within $1<Z<5\,{\rm kpc}$ is ${\rm d}\langle{\rm [Fe/H]}\rangle/{\rm d}Z\approx -0.19\,{\rm dex}\cdot{\rm kpc}^{-1}$ around $R=8.25\,{\rm kpc}$. But the mean radial gradient is almost negligible. The density profile of the thick disk is fitted with modified double exponential law decaying to a constant at far distance. The scale height and scale length thus estimated are $H\approx 1.13\,{\rm kpc}$ and $L\approx 3.63\,{\rm kpc}$, which are in consistent with the results determined from star-counts method in previous studies. The halos are described with two-axial power-law ellipsoid  and the axis ratios of both inner halo and outer halo, inferred from stellar number density in $R$-$Z$ plane, are $q_{ih}\approx 0.49$ and $q_{oh}\approx 0.61$, respectively. It also manifests that the outer halo is a more spherical than inner halo. Moreover, the halo power-law indices estimated are $n_{ih}\approx 3.4$ and $n_{oh}\approx 3.1$, indicating that the stellar number density of inner halo changes more steeper than that of outer halo.  
\end{abstract}

\keywords{Galaxy: disk -- Galaxy: halo -- Galaxy: abundances--Galaxy:fundamental parameters}

\section{Introduction}

\par The Milky Way Galaxy is presumably composed of several stellar components \citep[][and references therein]{Freeman2002, Juric2008, Ivezic2008, Bond2010, Ivezic2012, Bland2016}. The characterization of these stellar components needs measurements and analysis of the properties of large samples of individual stars in the phase space spanned by spatial coordinate, velocity components, and metallicity. With the photometric data from the Sloan Digital Sky Survey \citep[SDSS;][]{York2000}, the map of stellar number density can be constructed, which enables detailed investigation of the stellar populations in marginal space. However, the chemical map of the Galaxy has not been directly obtained, due to the limited sky and depth coverage of spectroscopic surveys, e.g. the Sloan Extension for Galactic Understanding and Exploration \citep[SEGUE, a subsurvey of SDSS;][]{Yanny2009}, the Radial Velocity Experiment \citep[RAVE;][]{Steinmetz2006}, and the Large Sky Area Multi-Object Fiber Spectroscopic Telescope \citep[LAMOST;][]{Cui2012, Deng2012, Zhao2012}. This motivates astronomers to find alternative way to obtain estimates of stellar metallicity. Because of the detectable effect of the exhaustion of metals in stellar atmosphere on the emergent flux \citep{Schwarzschild1955}, photometric method is widely adopted to derive estimates of stellar metallicity. The typical applications in this regard based on photometric data can be found in some studies \citep{Du2004, Karaali2005, Ivezic2008, Peng2012, Peng2013, An2013, An2015, Gu2015,Yuan2015, Tuncel2017}.

\par Despite its apparent advantage, photometric metallicity calibrations are relatively not as accurate as those from spectroscopic observations and sometimes yield poor results for very metal-rich or very metal-poor stars, this is mainly due to two reasons: one is the calibration methods themselves and the other is photometric errors. Most photometric metallicity calibrations are characterized by their assignment of stellar metallicity individually based on color indices. \cite{Gu2016a} presented a new method  to derive stellar metallicity distributions based on SDSS photometric data, taking advantage of the Monte-Carlo technique in order to statistically reduce the uncertainties in this estimation. Any metallicity estimate with SDSS photometric data is very sensitive to $u$-band magnitude due to its relatively large error, especially at faint end. This was alleviated by the advent of South Galactic Cap of the u-band Sky Survey (SCUSS), which re-surveyed the $u$-band magnitude in south Galactic Cap with $5$-min exposure time, resulting in $\sim 1.2$ magnitude deeper of SCUSS $u$-band magnitude than that of SDSS. With SCUSS $u$-band magnitude, \cite{Gu2016b} developed a method to statistically convert SDSS $u$-band magnitude to SCUSS $u$-band magnitude, during which the noise of SDSS $u$-band magnitude is damped, thus leading to high accuracy of SDSS $u$-band magnitude as same as SCUSS $u$-band magnitude.

\par In the present work, our purpose is to study the properties of Galactic populations jointly from both chemical and spatial distributions. We first convert SDSS $u$ to SCUSS $u$, and then, together with other magnitudes of SDSS, estimate the stellar photometric MDFs of large number of individual stars using the method developed in \cite{Gu2016a}. Photometric MDFs of stars at different location are estimated, from which we give the mean metallicity gradient. After that, we study the spatial properties of the Galactic populations separated from MDFs through Gaussian fitting. In particular, we focus on the scale length and scale height of the thick disk,  axis ratios and power-law indices of inner halo and outer halo.

\par This paper is organized as follows. We begin in Section 2 with a brief overview of the two methods developed in \cite{Gu2016a, Gu2016b}. In Section 3, we describe the method of selecting F/G main-sequence stars as samples. In Section 4, we estimate photometric MDFs by Monte-Carlo-based method and derive the vertical and radial gradients of mean metallicity . In Section 5, we introduce a procedure to separate Galactic stellar populations from MDFs, and the studies of density profiles of thick disk and halo(s) are followed in Section 6. The conclusions and perspectives are given in Section 7.

\section{Brief Overview of Gu's Methods}

\par In this section, we briefly review the two methods developed, which are all based on the Monte-Carlo technique. More detailed presentation of the methods could be found in \cite{Gu2016a, Gu2016b}.

\subsection{The Method for Estimating Photometric MDFs}

\par Most previous photometric metallicity calibrations are characterized by the one-to-one correspondence between stellar metallicity and color indices for an individual stars. However, in order to investigate the chemical structure of the Galactic stellar populations, we only require knowledge of the MDF for a large statistical sample of stars. In addition, metallicity of single star is actually uncertain even though its color indices are fixed, varying metallicity estimate to form a distribution. Enlightened by this fact, \cite{Gu2016a} obtained the statistics of stellar metallicities from a large number of spectroscopically surveyed stars with their both color indices $u-g$ and $g-r$ respectively in a fixed bin, and interpreted this statistics as a result of the intrinsic probability distribution of metallicity. The statistics is recorded in so-called  ``seed" array, with column associated with color indices, the row associated with metallicity, and elements holding the pertinent information of probability distribution. For a single star, \cite{Gu2016a} used the Monte-Carlo technique to generate a value according to the metallicity probability distribution corresponding to this star's color indices, and this value is associated as the metallicity of this star. Although a single star's metallicity obtained in this way is uncertain, the  metallicities of a large number of stars can derive a stable MDF from which we can extract the chemical properties of Galactic populations.

\subsection{The Method for improving the Accuracy of SDSS $u$-Band Magnitude}

\begin{figure}
\resizebox{1.0\hsize}{!}{\includegraphics{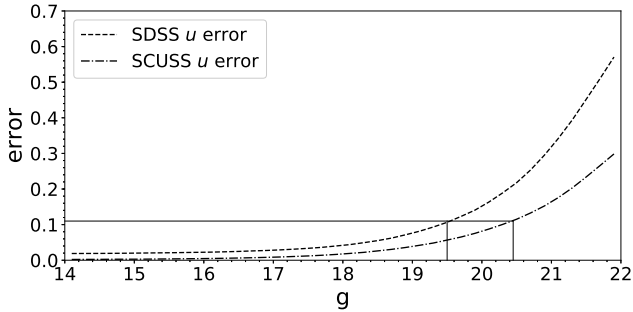}}
\caption{Average $u$ (SDSS and SCUSS) error as a function of $g$-band magnitude. Main-sequence stars with $0.2<g-r<0.8$ are selected. It is obvious that the error of the SDSS $u$ is much larger than that of the SCUSS $u$, especially at the faint end.}
\label{fig_error}
\end{figure}

\par The relatively larger error of SDSS $u$-band magnitude, especially at faint end, often impose limitation on the range of application of the photometry. For example, photometric metallicity estimator given by \cite{Ivezic2008} can only be applied for stars brighter than $g=19.5$. This is determined by the limiting magnitude of the spectroscopically surveyed stars, which set the maximum SDSS $u$-band magnitude error. South Galactic Cap of the $u$-band Sky Survey (SCUSS) also provides the $u$-band magnitude with much more accuracy for stars in south Galactic cap. There exists very minor differences between SDSS $u$ filter and SCUSS $u$ filter, with the response curve of SCUSS $u$ filter slightly narrower and having 24 \AA blueshift. Thus, we can reasonably make an equivalence between these two bands, as already demonstrated by \cite{Gu2016b}. Figure \ref{fig_error} shows the average $u$ error of both SDSS and SCUSS versus $g$-band magnitude for main-sequence stars with $0.2<g-r<0.8$. Apparently, SCUSS $u$ is more accurate than SDSS $u$. Here, it is noted that the SCUSS $u$ error when $g=20.5$ is equal to the SDSS $u$ error when $g=19.5$. Thus, we can infer that the photometric metallicity estimator based on SCUSS $u$ can be safely applied up to $g=20.5$, one magnitude deeper than earlier. After position matching, we can obtain a merged stellar catalog, in which each star has both magnitudes of SDSS $u$-band and SCUSS $u$-band.   

\par
\begin{figure}
\begin{center}
\begin{minipage}[t]{1.00\hsize}
\resizebox{1.0\hsize}{!}{\includegraphics{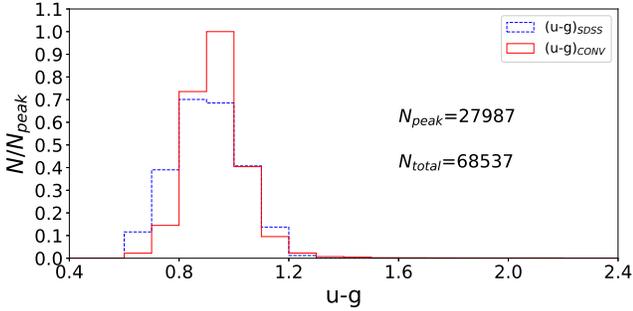}}
\end{minipage}
\end{center}
\caption{The histograms of $(u-g)_{\rm SDSS}$ and $(u-g)_{\rm CONV}$, which are respectively the colors from SDSS photometric data and their converted ones. These stars  are randomly selected from SDSS catalogue with the conditions $0.2<g-r<0.3$ and $20.0<g<20.5$.}
\label{fig_histgram_comparison}
\end{figure}

 For all of the main-sequence stars with $g$ magnitude and color index $g-r$ fixed in a small bin, \cite{Gu2016b} obtained a statistics and recorded the result in a so-called ``convertor" array, with the column associated with color index $(u-g)_{\rm SDSS}$, the row associated with color index $(u-g)_{\rm SCUSS}$, the elements holding the counts of stars. The statistics recorded in such ``convertor" arrays is interpreted as the probability distribution due to the errors of both SDSS $u$ and SCUSS $u$. For a single star with given $(u-g)_{\rm SDSS}$ and no $(u-g)_{\rm SCUSS}$, Gu used the Monte-Carlo technique to generate a value according to the probability distribution of $(u-g)_{\rm SCUSS}$ corresponding to this given $(u-g)_{\rm SDSS}$, and consider this value as the new $u-g$, which can be understood as the converted $(u-g)_{\rm SCUSS}$ of this star. Although a single star's $u-g$ thus obtained is uncertain, the indices of a large number of stars lead to a statistically stable distribution which is narrower than that previous conversion. This is mainly due to that SCUSS $u$ is more accurate than SDSS $u$. In Figure \ref{fig_histgram_comparison}, we compare the dispersions between $(u-g)_{\rm SDSS}$ colors and their converted $(u-g)_{\rm CONV}$. These stars are randomly selected from SDSS catalogue, not those used to construct ``convertor" arrays. From the comparison between two histograms, we may convince that the conversion of $(u-g)_{\rm SDSS}$ indeed reduces the error. In the following sections, the subscripts specifying $u$-band magnitude are no longer needed, as they are considered as the converted magnitudes with their improved accuracy.

\par More detailed information and data reduction about SCUSS, please refer to \cite{Zhou2016} and \cite{Zou2015, Zou2016}. The public SCUSS data can be accessed from the official website \emph{http://batc.bao.ac.cn/Uband/}.

\section{Selection of F/G Main-sequence Stars As Samples}

\par In this section, we select the main-sequence stars by rejecting those objects far from stellar locus, and detailed procedure was given in \cite{Jia2014}.  
We further impose the criteria as follows:
\begin{itemize}
\item $14<g<20.5$,
\item $0.2<g-r<0.4$,
\item $0.6<u-g<2.2$,
\end{itemize}
with the first one ensuring that the converted $u$ magnitudes of photometrically surveyed stars are as accurate as those of spectroscopically surveyed stars, and the subsequent two selecting the F/G main-sequence stars. The above selecting criteria are required just to make the sample stars suitable for the estimation of photometric MDFs. Note that the magnitudes throughout this paper have already been corrected for extinction using the values from \cite{SchlaflyFinkbeiner2011}.

\par We calculate the distances to our sample stars with SDSS $r$-band absolute magnitudes estimated from the photometric parallax relation
\begin{align}
M_r= & 3.2+13.30(r-i)-11.50(r-i)^2 \nonumber \\
& +5.40(r-i)^3-0.70(r-i)^4 \text{,}
\end{align}
which is provided by \cite{Juric2008}. Then, combining the distance $D$, Galactic longitude $b$ and latitude $l$, the position of each star can be determined, thus leading to three-dimensional map construction of the sample stars. {It should be noted that the derivation of the above photometric parallax relation was based on the old extinction map \citep{Schlegel1998}. However, we assume that the difference of the extinction maps is too small and can be safely neglected.  Assuming the Galaxy is axis-symmetric, the cylindrical galactocentric coordinate system $(R,\,Z,\,\phi)$ is adopted for convenience, which is obtained through the following set of coordinate transformations
\begin{align}
&X=R_{\circleddot}-D\cos(l)\cos(b) \text{,} \\
&Y=-D\sin(l)\cos(b) \text{,} \\
&R=\sqrt{X^2+Y^2} \text{,} \\
&Z=D\sin(b) \text{,} \\
&\phi=\arctan(Y/X) \text{,}
\end{align}
where $R_{\circleddot}=8\,{\rm kpc}$ is the adopted distance of the Sun to the Galactic center \citep{Reid1993,Bland2016}.

\begin{figure}
\begin{center}
\begin{minipage}[t]{1.00\hsize}
\resizebox{1.0\hsize}{!}{\includegraphics{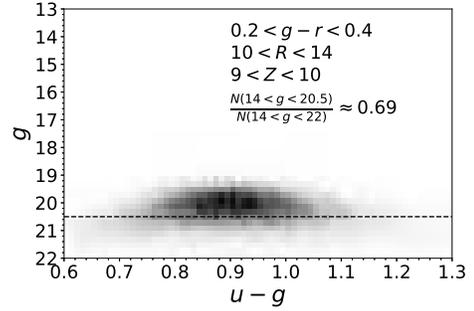}}
\end{minipage}
\end{center}
\caption{Color-magnitude diagram, $g$ vs $u-g$. The stars are selected within $0.2<g-r<0.4$, $10<R<14\,{\rm kpc}$ and $9<Z<10\,{\rm kpc}$, as also labeled. The ratio of number of stars confined with $14<g<20.5$ over that with $14<g<22$ is approximately $0.69$.}
\label{fig_CMD}
\end{figure}

\par In Figure \ref{fig_CMD}, we present the color-magnitude diagram, $g$ vs $u-g$ with the sample stars in the region within $10<R<14\,{\rm kpc}$ and $9<Z<10\,{\rm kpc}$, which are relatively distant. In this region, the number of F/G stars with $14<g<20.5$ accounts up to 69\% of number of those with $14<g<22$. It can be clearly seen that if there is no improvement of $u$ magnitude accuracy, we can only select the F/G stars with $14<g<19.5$ which is a much smaller fraction than the total F/G stars. Thus, it increases dramatically the fraction of distant stars, demonstrating the merit of improving $u$ magnitude error with the second method briefly reviewed in the previous section.

\section{Vertical and Radial Gradients of Mean Metallicity}

\begin{figure*}
\begin{center}
\begin{minipage}[t]{0.49\hsize}
\resizebox{1.0\hsize}{!}{\includegraphics{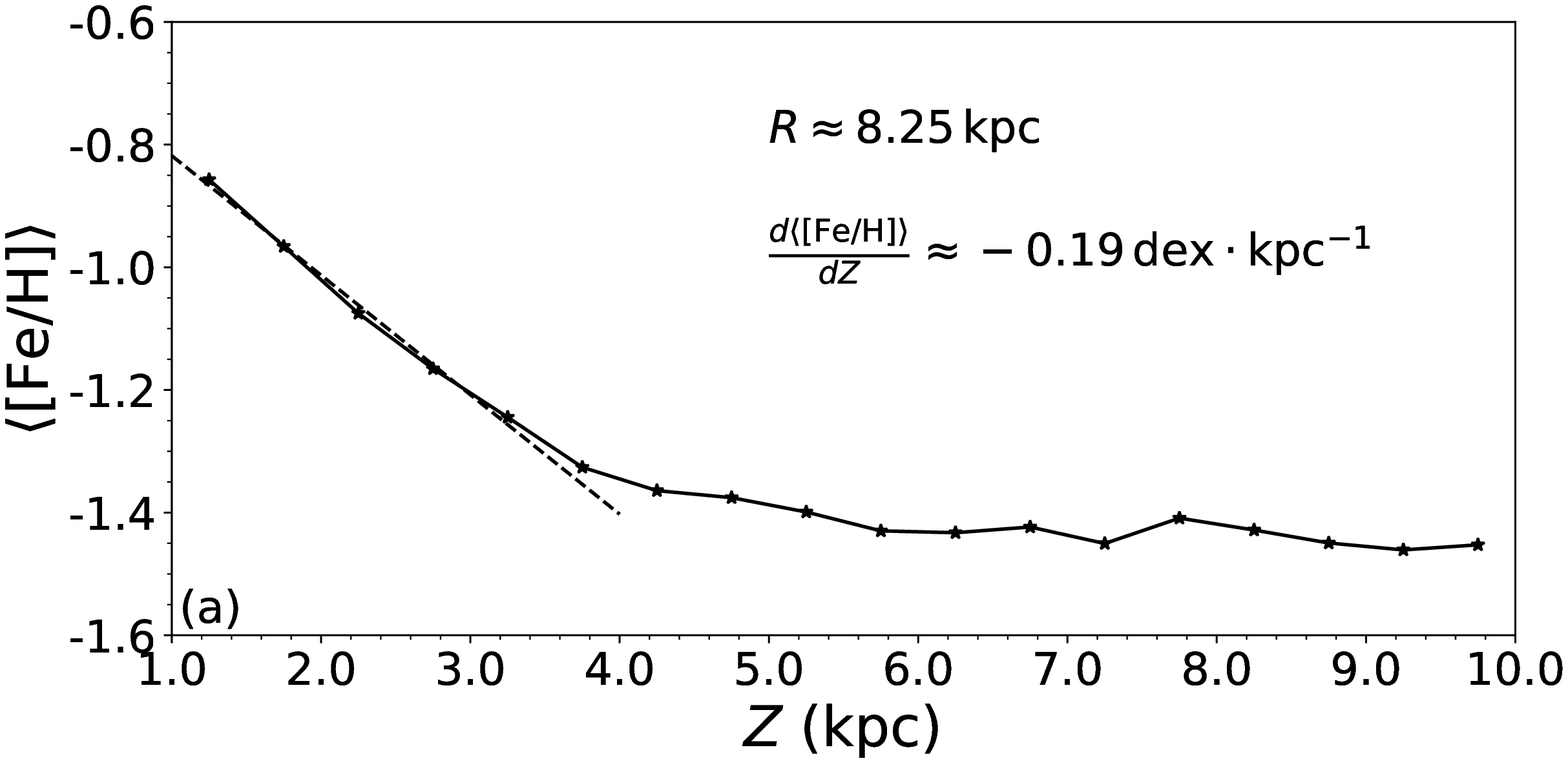}}
\end{minipage}
\begin{minipage}[t]{0.49\hsize}
\resizebox{1.0\hsize}{!}{\includegraphics{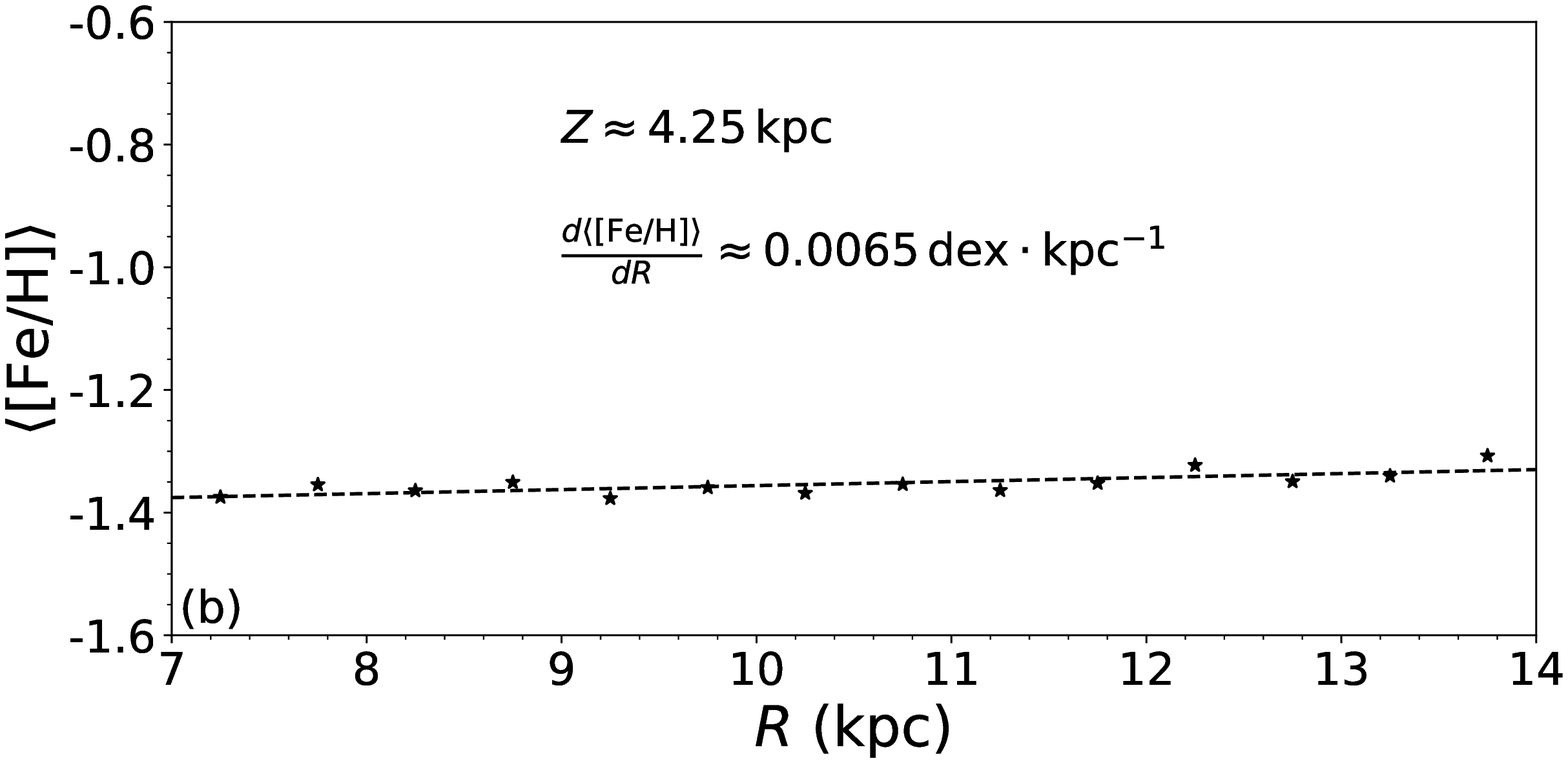}}
\end{minipage}
\end{center}
\caption{Panel (a): The mean metallicity $\langle{\rm [Fe/H]}\rangle$ as a function of $Z$, with $R$ being around $8.25\,{\rm kpc}$. The straight dashed line fits the data points within $1\,{\rm kpc}<Z<5\,{\rm kpc}$, with slope about $-0.19$. Panel (b): The mean metallicity $\langle{\rm [Fe/H]}\rangle$ as a function of $R$, with $Z$ being around $4.25\,{\rm kpc}$. The data points are fitted with a straight line whose slope is negligible.}
\label{fig_metallicity_gradients}
\end{figure*}

\par In this section, we use Monte-Carlo-based method to estimate the photometric MDFs for the sample stars located in certain three-dimensional spatial region. Particularly, we focus on the how MDFs vary with spatial position. As examples, we present four photometric MDFs for different spatial regions in Figure \ref{fig_MDF}, and the bimodality in each MDF is easily seen. For each MDF $f$, we can define the mean metallicity as follows:
\begin{align}
\langle{\rm [Fe/H]\rangle}\equiv \int xf(x){\rm d}x \text{,}
\end{align}
where $x\equiv {\rm [Fe/H]}$. When MDF is in the form of histogram, the integral should be replaced with summation, just as the case in this paper. In the Figure \ref{fig_metallicity_gradients}(a), the spatial dependence of mean metallicity $\langle{\rm [Fe/H]}\rangle$ along vertical direction around $R=8.25$ is clearly seen. Within $1<Z<5\,{\rm kpc}$, the negative gradient of $\langle{\rm [Fe/H]}\rangle$ can be derived by a fitted straight line,
\begin{align}
\frac{{\rm d}\langle{\rm [Fe/H]}\rangle}{{\rm d}Z}\approx -0.19\,{\rm dex}\cdot{\rm kpc}^{-1} \text{.}
\end{align}
This value lies in the typical range $[0.0,\,-0.22]$ from previous works \citep{AllendePrieto2006, Chen2011, Bilir2012, Mikolaitis2014, Li2017, Duong2018}. Result from the Radial Velocity Experiment \citep[RAVE;][]{Steinmetz2006} also lie in this range \citep{Kordopatis2011, Ruchti2011, Katz2011, Carrell2012}. The negative vertical gradient can be explained that the relatively more metal-poor halo stars become dominating far from the Galactic plane. Similarly, the spatial dependence of $\langle{\rm [Fe/H]}\rangle$ along radial direction around $Z=4.25$ is studied, and no apparent gradient of $\langle{\rm [Fe/H]}\rangle$ is found, as shown in the Figure \ref{fig_metallicity_gradients}(b). This agrees with the previous works \citep{AllendePrieto2006, Nordstrom2004, Ruchti2011, Bilir2012, Coskunoglu2012}.

\section{Galactic Stellar Populations from Photometric MDFs}

\begin{figure}
\begin{center}
\begin{minipage}[t]{1.00\hsize}
\resizebox{1.0\hsize}{!}{\includegraphics{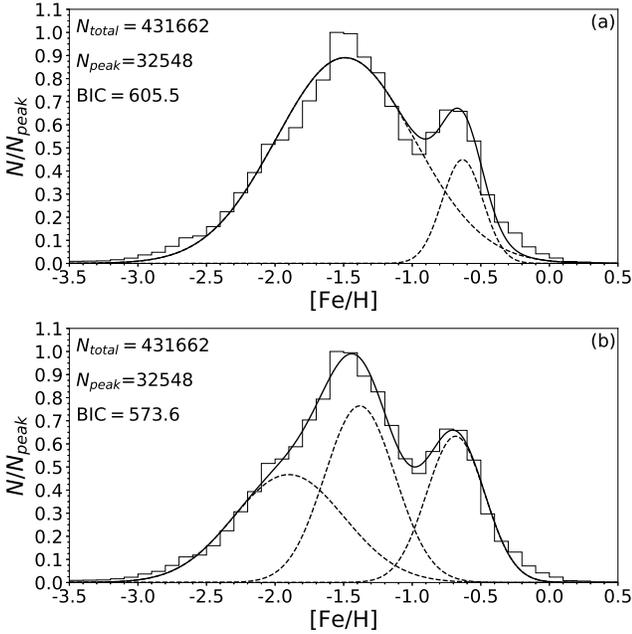}}
\end{minipage}
\end{center}
\caption{ Photometric MDFs of the F/G main-sequence stars $0.2<g-r<0.4$. The selected stars are in the certain spatial blocks listed in Table \ref{tab_blocks}. The MDFs are fitted with two-/three-Gaussian model (upper/lower panel). The histograms in both panels are the same. In each panel, the peak value of histogram is normalized to one, with the actual values labeled. The values of Bayesian information criterion (BIC) for both fits are shown. The best-fit parameters of the three-Gaussian model are listed in Table \ref{tab_gaussian_parameters}.}
\label{fig_MDF_model}
\end{figure}

\begin{table}
\caption{The best-fit values of mean and variance of each metallicity distribution of Gaussian form.}
\begin{center}
\begin{tabular}{>{\centering\arraybackslash}m{0.5cm}|>{\centering\arraybackslash}m{1.6cm}|>{\centering\arraybackslash}m{1.6cm}|>{\centering\arraybackslash}m{1.6cm}}
\hline
\hline
 & thick disk & inner halo & outer halo \bigstrut \\ \hline
$\mu$  & $-0.68\pm 0.014$ & $-1.38\pm 0.028$ & $-1.90\pm 0.19$ \bigstrut \\ \hline
$\sigma$  & $0.21\pm 0.012$ & $0.26\pm 0.035$ & $0.40\pm 0.09$ \bigstrut \\ \hline
\hline
\end{tabular}
\label{tab_gaussian_parameters}
\end{center}
\end{table}

\begin{figure*}
\begin{center}
\begin{minipage}[t]{1.00\hsize}
\resizebox{1.0\hsize}{!}{\includegraphics{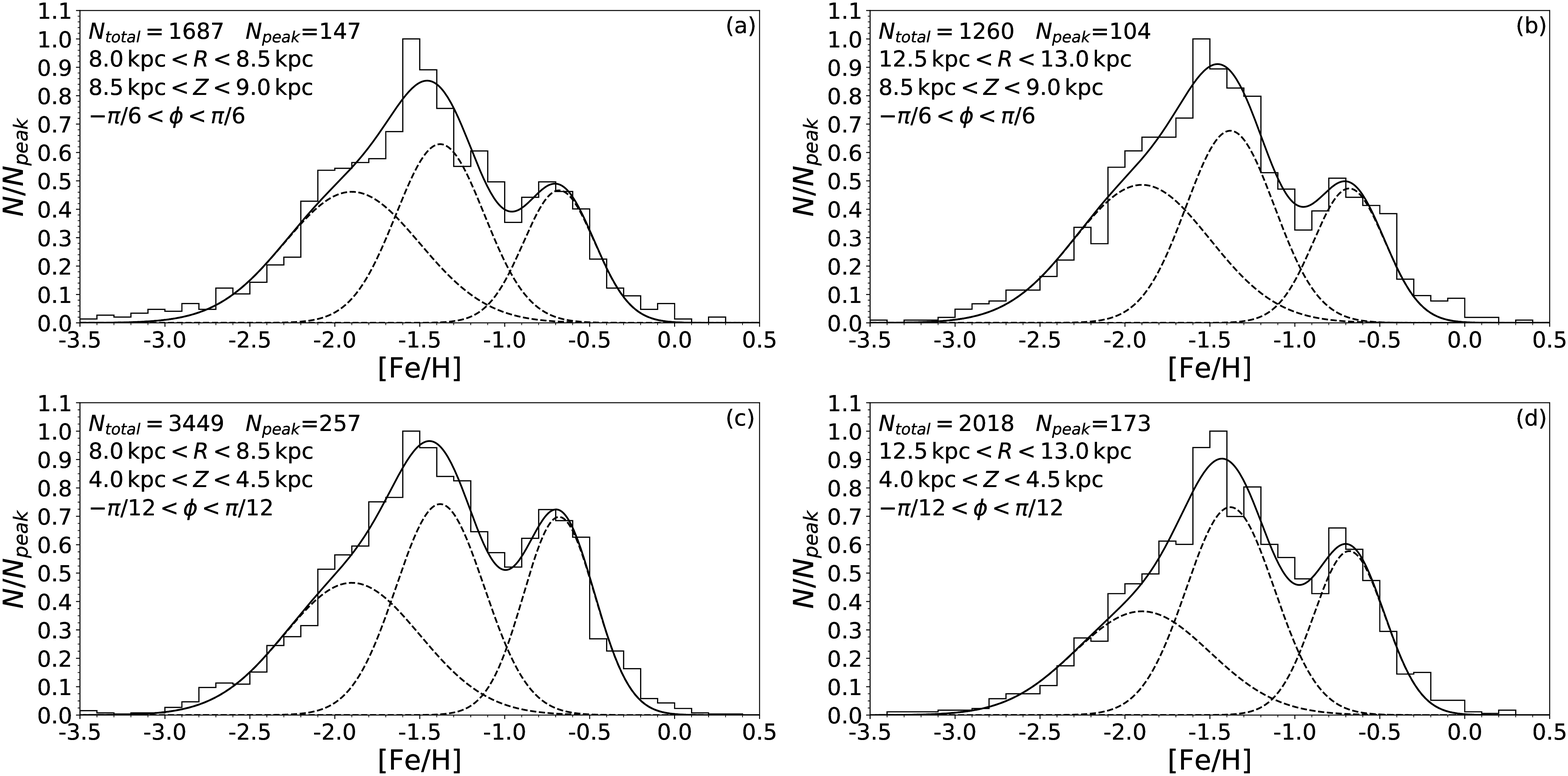}}
\end{minipage}
\end{center}
\caption{Photometric MDFs of the F/G main-sequence stars with $0.2<g-r<0.4$. Besides, the stars are confined in certain spatial region, as labeled in each panel. Each MDF is fitted with three gaussians. The mean and variance of each Gaussian component are fixed, with values listed in Table \ref{tab_gaussian_parameters}. The peak values of these four histograms are all normalized to one, with actual values also labeled, together with the value of the total number of stars.}
\label{fig_MDF}
\end{figure*}

\par We make the assumption that photometric MDFs of stars from a single stellar population are well described by a single Gaussian function. To describe the gross behavior of photometric MDF, we confine ourselves to the candidates of two-Gaussian model and three-Gaussian model. In Figure \ref{fig_MDF_model}, we show the photometric MDFs for large number of sample stars locating in spatial blocks listed in Table \ref{tab_blocks}, and respectively fit the MDFs with the two candidate models. For each fit, we calculate the goodness-of-fit statistics according to
\begin{align}
{\rm BIC}=N\ln(\chi^2/N)+\ln(N)N_{\rm varys} \text{,}
\end{align}
which is widely known as Bayesian information criterion (BIC) \citep{Ivezic2014}. Here, $N$ is the number of data points, $N_{\rm varys}$ is number of variable parameters, and $\chi^2$ is defined as
\begin{align}
\chi^2=\sum_i^N r_i^2 \text{,}
\end{align}
where $r=(data-model)/uncertainty$\footnote{We assume the uncertainties of all data points to be one.}. Since the BIC is smaller for the three-Gaussian fit than for the two-Gaussian fit (shown as legends in Figure \ref{fig_MDF_model}), we adopt the three-Gaussian model to fit photometric MDF. With Bayesian information criterion (BIC), \cite{Zuo2017} also determined that the optimal number of Gaussians describing Photometric MDFs is three. Table \ref{tab_gaussian_parameters} list the characteristic parameters of the three Gaussians determined from the fit shown in Figure \ref{fig_MDF_model}(b). It needs to be noted here that we have excluded the abnormal combinations of Gaussians by assigning parameters the corresponding ranges when performing the three-Gaussian fit. In the following model-fit of photometric MDFs for the sample stars in the sub-regions of spatial blocks listed in Table \ref{tab_blocks}, we fix means and variances of three Gaussians, with the values taken from Table \ref{tab_gaussian_parameters}, and leaving the weights of each Gaussian to be variables (see Figure \ref{fig_MDF}). Here, we focus on the statistical weight of each Gaussian component. In Figure \ref{fig_MDF}, the three Gaussians are associated with the thick disk, inner halo, and outer halo, respectively. We ignore the thin disk  due to few contribution in star counts above $3\,{\rm kpc}$. Here, we adopt the inner-/outer-halo dichotomy, which means that the halo is composed of two overlapping components. Many studies have provided evidences that these two halos have distinct spatial density profiles, kinematics, and metallicities \citep[e.g.][]{Carollo2007, Carollo2010, Carollo2012, Carollo2014, An2013, An2015, Zuo2017, Liu2018}.

\par Before the thick disk and both halos can be studied in detail, we develop a statistically robust scheme for classifying stars into three components. We describe MDF $f$ for each cell using the function as follows:
\begin{align}
f=f_{td}+f_{ih}+f_{oh}=\sum_{a=td,ih,oh}\frac{A_a}{\sqrt{2\pi}\sigma_a}\exp\left[-\frac{(x-\mu_a)^2}{2\sigma_a^2}\right] \text{,}
\end{align}
where $x\equiv{\rm [Fe/H]}$. Obviously, $f$ is the sum of three Gaussians with different weights $A_{th}$, $A_{ih}$, and $A_{oh}$. Denoting $N_{total}$ as the total number of sample stars in a certain spatial region, the numbers of sample stars belonging to each component are calculated as follows:
\begin{align}
& N_{td}=N_{total}\cdot A_{td}/(A_{td}+A_{ih}+A_{oh}) \text{,} \\
& N_{ih}=N_{total}\cdot A_{ih}/(A_{td}+A_{ih}+A_{oh}) \text{,} \\
& N_{oh}=N_{total}\cdot A_{oh}/(A_{td}+A_{ih}+A_{oh}) \text{.}
\end{align}
Thus, the Galactic stellar populations are separated successfully. Despite the superiority of three-Gaussian model, we still caution that the two-Gaussian model seems also good. The three-Gaussian model of photometric MDFs just represents a classification scheme, from which we hope to extract more detailed information about Galactic structures.

\section{Spatial Structures of Thick Disk and Halo(s)}

\begin{table*}
\caption{The boundaries of spatial blocks in which the sample stars are selected.}
\begin{center}
\begin{tabular}{>{\centering\arraybackslash}m{0.5cm}|>{\centering\arraybackslash}m{1.5cm}>{\centering\arraybackslash}m{1.5cm}>{\centering\arraybackslash}m{1.5cm}>{\centering\arraybackslash}m{1.5cm}>{\centering\arraybackslash}m{1.5cm}>{\centering\arraybackslash}m{1.5cm}}
\hline
\hline
id & $R_{min}$ (kpc) & $R_{max}$ (kpc) & $Z_{min}$ (kpc) & $Z_{max}$ (kpc) & $\phi_{min}$ (rad) & $\phi_{max}$ (rad) \bigstrut \\ \hline
1 & $7.0$ & $14.0$ & $8.0$ & $10.0$ & $-\pi/6$ & $\pi/6$ \bigstrut \\ \hline
2 & $7.0$ & $14.0$ & $5.0$ & $8.0$ & $-\pi/12$ & $\pi/6$ \bigstrut \\ \hline
3 & $7.0$ & $14.0$ & $4.0$ & $5.0$ & $-\pi/12$ & $\pi/12$ \bigstrut \\ \hline
4 & $7.0$ & $14.0$ & $3.0$ & $4.0$ & $-\pi/24$ & $\pi/12$ \bigstrut \\ \hline
\hline
\end{tabular}
\end{center}
\label{tab_blocks}
\end{table*}

\par We only consider the stars in the spatial blocks listed in Table \ref{tab_blocks}. The range of azimuthal angle for each block is determined by the available surveyed stars (see Figure \ref{fig_spatial_region}). In $R$-$Z$ plane, these spatial blocks are confined in the rectangle region
\begin{align}
7<R<14\,{\rm kpc} \hspace{0.5cm}\text{and}\hspace{0.5cm} 3<Z<10\,{\rm kpc} \text{.} \label{eq_RZ_region}
\end{align}
We bin the blocks in this plane into $0.5\,{\rm kpc}\times 0.5\,{\rm kpc}$ cells which are further extended to form a three-dimensional (3D) cells combined with finite range of azimuthal angle. We count the number of sample stars in each 3D cell. Following the procedure described in the previous section, the number of stars belonging to each component in each 3D cell can be computed. Divided by the volume of each 3D cell, stellar numbers become stellar number densities, thus leading to construction of stellar number density map for each component in the considered region. It is here noted that the rectangle region is deliberately chosen to avoid any already observed overdensities, e.g. Virgo and Monoceros, whose locations are indicated in Figure \ref{fig_spatial_region}. In the following, the stellar number density maps of thick disk, inner halo, and outer halo are subjecting to make model-fitting\footnote{The routine \textit{Model} provided by Python package \textit{Lmfit} is a very handy for curve fitting.} based on Levenberg-Marquadt minimization algorithm \citep{Press2007}.

\begin{figure*}
\begin{center}
\begin{minipage}[t]{0.99\hsize}
\resizebox{1.0\hsize}{!}{\includegraphics{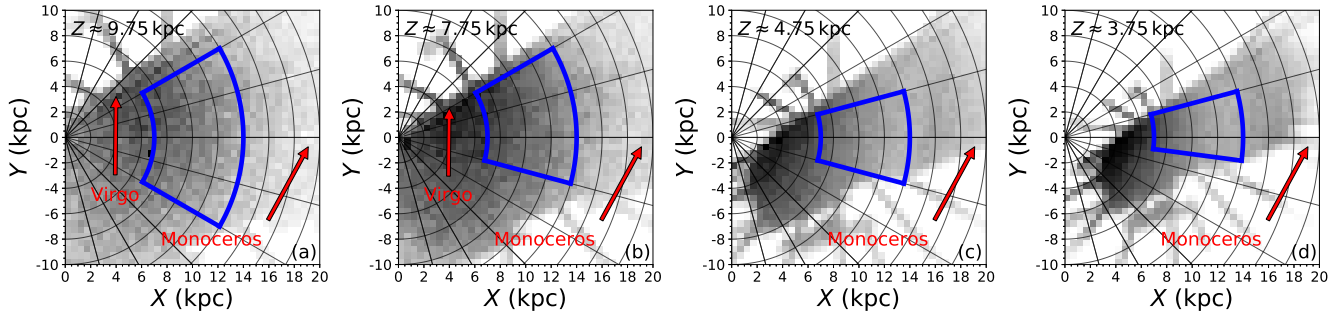}}
\end{minipage}
\end{center}
\caption{The slices of density profile in $X$-$Y$ plane at four different heights $Z$ (shown in each plane). The sector areas circled with thick blue lines in four planes correspond to four spatial blocks listed in Table \ref{tab_blocks}. The arrows indicate the direction of Virgo and Monoceros, which are however not evident in these panels.}
\label{fig_spatial_region}
\end{figure*}

\subsection{Scale Height and Scale Length of the Thick Disk}

\par We describe the thick disk in $R$-$Z$ plane with a modified double exponential function
\begin{align}
\rho_D=A_D\exp\left[-\frac{Z}{H}-\frac{R}{L}\right]+C_D \text{,} \label{eq_disk_model}
\end{align}
where $H$ and $L$ are the scale height and scale length. Here, $A_D$ and $C_D$ are parameters, with the latter being added due to the observation that the thick disk densities in the concerned region do not show the trend of decaying to zero. With the disk model shown in Eq. (\ref{eq_disk_model}), the best-fit values of scale height and scale length of thick disk are obtained, respectively being
\begin{align}
& H_{td}\approx 1.134\pm 0.027\,{\rm kpc} \text{,} \\
& L_{td}\approx 3.627\pm 0.088\,{\rm kpc} \text{,}
\end{align}

\par Recent literature (after 2000) reports thick disk scale length of 2-5 kpc, while the
thick disk scale height has reported values of about 500-1400 pc \citep{Chen2001, Siegel2002, Du2003,
Du2006, Larsen2003, Cabrera2005, Karaali2007, Juric2008, Yaz2010,Chang2011, Jia2014, Chen2017, Wan2017} .
There are also debates about the relative size of scale
length of the thick disk, as photometric
stellar density distribution generates longer scale length, while spectroscopic sample,
which usually defines the thick disk in abundance or age \citep{Xiang2018}, yielding shorter scale length for the thick disk 
\citep[e.g.][]{Bovy2012, Bovy2016, Cheng2012, Mackereth2017}. 
 Our results are generally in agreement with some photometric results.
Thus, we suggest that $C_D$ may be the result of the existence of metal-rich stars from Galactic halo. If so, the assumption made at the beginning of the previous section is not valid as expected. Some studies have also reported metal-rich stars in the halo based on spectroscopy \citep[e.g.][]{Ryan1991, Carlin2016}.

\begin{figure*}
\begin{center}
\begin{minipage}[t]{0.49\hsize}
\resizebox{1.0\hsize}{!}{\includegraphics{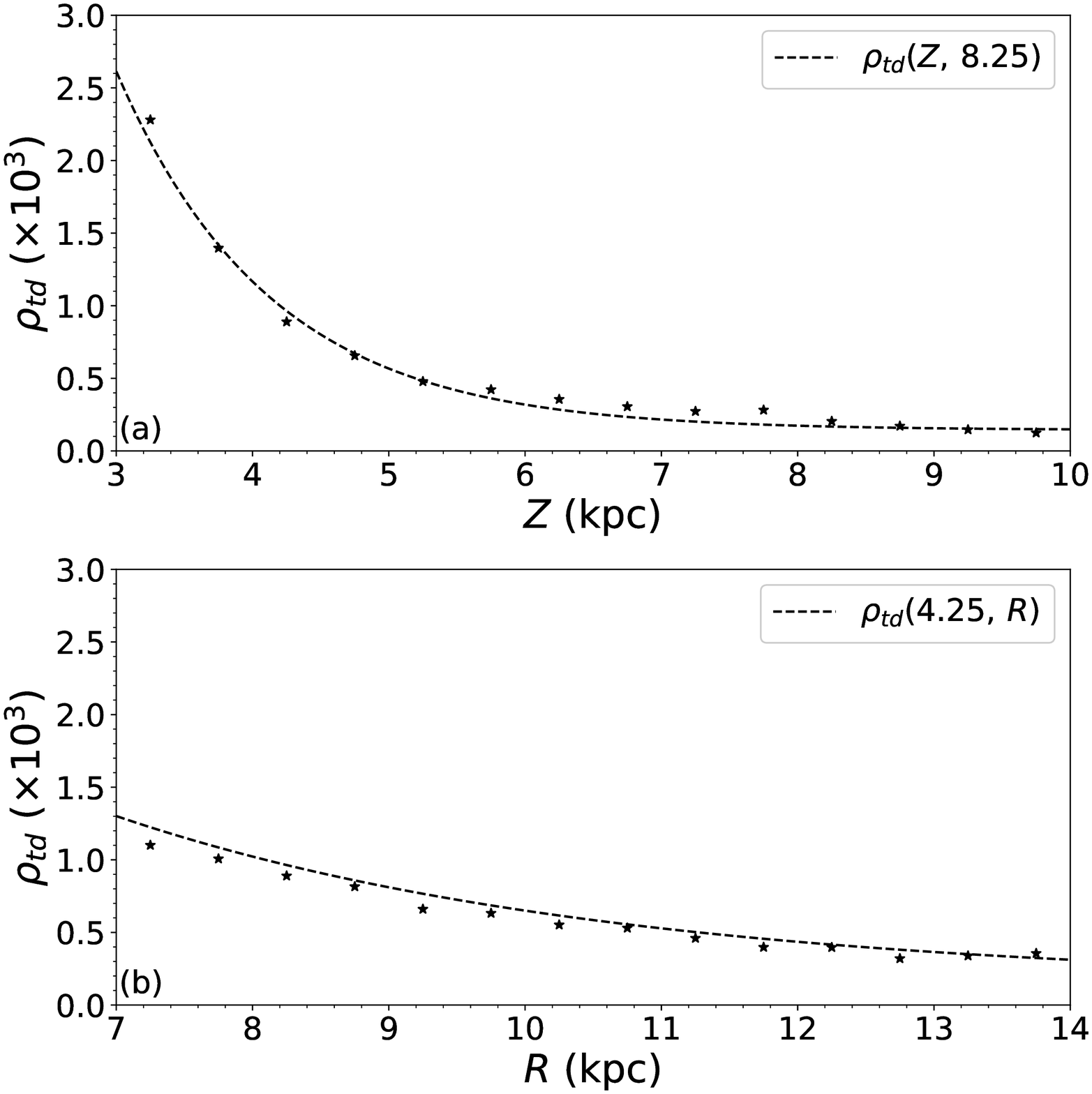}}
\end{minipage}
\begin{minipage}[t]{0.49\hsize}
\resizebox{1.0\hsize}{!}{\includegraphics{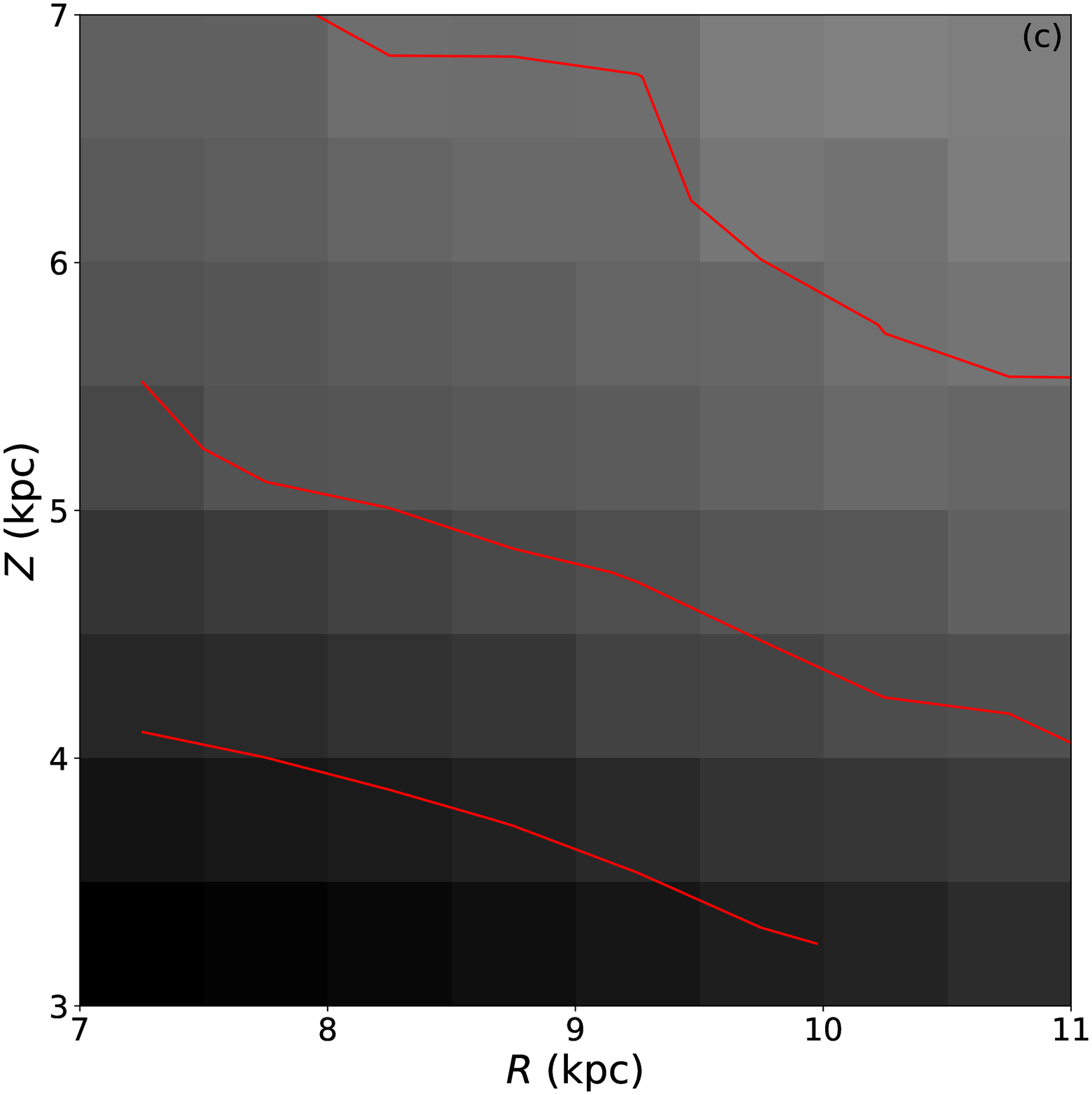}}
\end{minipage}
\end{center}
\caption{Panel (a): The density profile of thick disk stars along $Z$ direction around $R=8.25\,{\rm kpc}$. Panel (b): The density profile of thick disk stars along $R$ direction around $Z=4.25\,{\rm kpc}$. In each panel of these two, asterisks are from the data, dash line are generated from exponential function obtain from fitting. Panel (c): The map of $\ln(\rho_{td}-C_{td})$ as a function of $Z$ and $R$. Here, $\rho_{td}$ is the observed density of thick disk, and $C_{td}$ denoted the best-fit value of the constant $C_D$. Three contours are plotted.}
\label{fig_thick_disk_map}
\end{figure*}

\par In Figure \ref{fig_thick_disk_map}(a)(b), we show both vertical and radial slices of the stellar number density map of the thick disk. The asterisks are the observed densities, and the dash line along each direction is from the disk model with best-fit values of parameters.  It shows that the model generally fits the data well. According to Eq. (\ref{eq_disk_model}), we get the disk model in the following linear form
\begin{align}
\ln(\rho_D-C_D)=-\frac{Z}{H}-\frac{R}{L}+\ln(A_D) \text{,} \label{eq_disk_model_linear}
\end{align}
which is a two-variable linear function of $Z$ and $R$. The contours of the map from above function should be equi-spaced parallel lines. In Figure \ref{fig_thick_disk_map}(c), we draw the map of $\ln(\rho_{td}-C_{td})$, where $\rho_{td}$ is the observed density of thick disk, and $C_{td}$ is the best-fit value of parameter $C_D$. Three contours of this map are plotted, and they are founded to be a little bit deviated from the equi-spaced parallel lines. We consider this deviation arising from the errors and the difference between the observed data and best-fit model.

\subsection{Oblateness and Steepness of the Halos}

\par Evidence for the dual halo (the inner-halo and outer-halo populations) has been found by \cite{Carollo2007,Carollo2010} and \cite{Beers2012}.  Following works \citep[][]{deJong10, Kinman12, Kafle13, Hattori13, Chen14, Fernandez-Alvar15, Das16, Kafle17} that trace the more distant halo with giant stars or BHB stars also provide evidence for the duality of the halo. 
In general, the spatial structure of each halo component in $R$-$Z$ plane is typically described by a power-law ellipse,
\begin{align}
\rho_H=\rho_{\circleddot}\left[\frac{R_{\circleddot}}{\sqrt{R^2+(Z/q)^2}}\right]^{n} \text{,} \label{eq_halo_model}
\end{align}
where $\rho_{\circleddot}$ is the local stellar number density of halo, $q$ controls the ellipticity, and $n$ is the power-law index which indicates how fast stellar number density decrease with distance. In Figure \ref{fig_halo_maps}, the stellar number density maps in $R$-$Z$ plane for the inner halo and outer halo are given. For convenience, a quantity $r$ called radius is defined such that
\begin{align}
r^2\equiv R^2+\left(Z/q\right)^2 \text{,} \label{eq_radius}
\end{align}
which is the ellipse equation in $R$-$Z$ plane when radius $r$ takes a value. As implied by the Eq. (\ref{eq_halo_model}), we know that the contours of stellar number density from the halo can be approximated with ellipses described by Eq. (\ref{eq_radius}) when radius $r$ taking different values. Following this idea, we employ a two-stage procedure to estimate the parameters $q$ and $n$ for each halo. We firstly introduce an algorithm to determine the ellipses approximating the contours of density maps in Figure \ref{fig_halo_maps}, so that we can obtain the parameter $q$. Then, using the estimated value of $q$, we obtain the best-fit value of power-law index $n$ through linear regression with the following linear model
\begin{align}
\ln(\rho_{H})=-n\times\ln(r)+n\times\ln(R_{\circleddot})+\ln(\rho_{\circleddot}) \text{,} \label{eq_halo_model_linear}
\end{align}
which is developed from Eq. (\ref{eq_halo_model}).

\begin{figure*}
\begin{center}
\begin{minipage}[t]{1.00\hsize}
\resizebox{1.0\hsize}{!}{\includegraphics{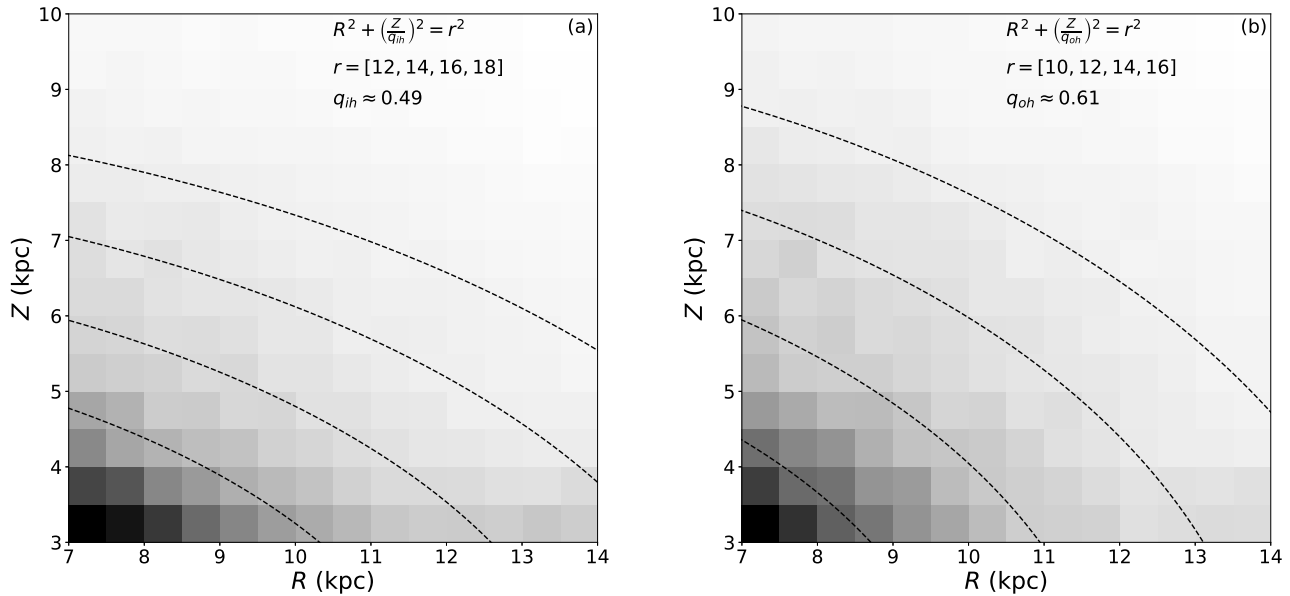}}
\end{minipage}
\end{center}
\caption{Panel (a): The density map of inner halo. Panel (b): The density map of outer halo. In each panel, several ellipses with same ellipticity approximating the contours are plotted, with function forms and the value of parameter $q$ controlling ellipticity labeled.}
\label{fig_halo_maps}
\end{figure*}

\par We now introduce the algorithm in a general way. Assume that we are given a rectangle density map confined by $R_{min}<R<R_{max}$ and $Z_{min}<Z<Z_{max}$, and it can be discretized into cells with $\Delta R$ and $\Delta Z$. For each cell, we have the density $\rho_a$ and the corresponding position $(R_a,\,Z_a)$, where the subscript $a$ denotes the index in any way. Given the parameter $q$ with a specific value, we can derive $r_a$ for each cell according to the definition of radius by Eq. (\ref{eq_radius}). Correspondingly, the boundaries of radius $r_{min}$ and $r_{max}$ are defined as $R_{min}^2+(Z_{min}/q)^2$ and $R_{max}^2+(Z_{max}/q)^2$. The whole interval $[r_{min}, r_{max}]$ is divided into $n$ equi-spaced intervals delimited with nodes $\{q_0=q_{min},\cdots, q_j,\cdots,q_n=q_{max}\}$. According to which interval each cell's radius $r_a$ falls in, we can group the cells. $\rho_a^{(j)}$s are the densities for those cells whose radii $r_a$s all satisfy $r_{j-1}<r_a<r_j$. For the densities $\rho_a^{(j)}$s with different index $a$, we can further define their variance
\begin{align}
var_j=\sum_a\left[\rho_a^{(j)}\right]^2-\left[\sum_a\rho_a^{(j)}\right]^2 \text{.}
\end{align}
Finally, we reach the quantity defined by
\begin{align}
var_{sum}=\sum_jvar_j \text{,}
\end{align}
which can also be seen as a function of $q$ in a implicit way. The best-fit value of $q$ can be derived when $var_{sum}$ takes the minimum value.

\begin{figure*}
\begin{center}
\begin{minipage}[t]{1.00\hsize}
\resizebox{1.0\hsize}{!}{\includegraphics{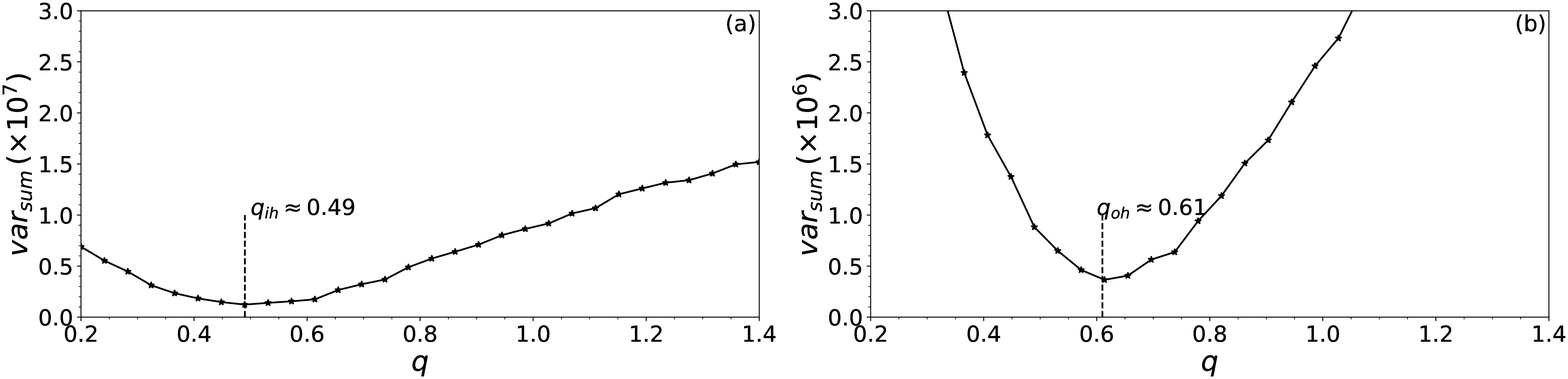}}
\end{minipage}
\end{center}
\caption{The defined quantity $var_{sum}$ versus $q$, from which the halo ellipticity can be determined when the function takes minimum value. Panel (a) is for inner halo, while the panel (b) is for outer halo. The parameters controlling ellipticity for both halos are $q_{ih}\approx 0.49$ and $q_{oh}\approx 0.61$, as also labeled in panels.}
\label{fig_q}
\end{figure*}

\begin{figure*}
\begin{center}
\begin{minipage}[t]{1.00\hsize}
\resizebox{1.0\hsize}{!}{\includegraphics{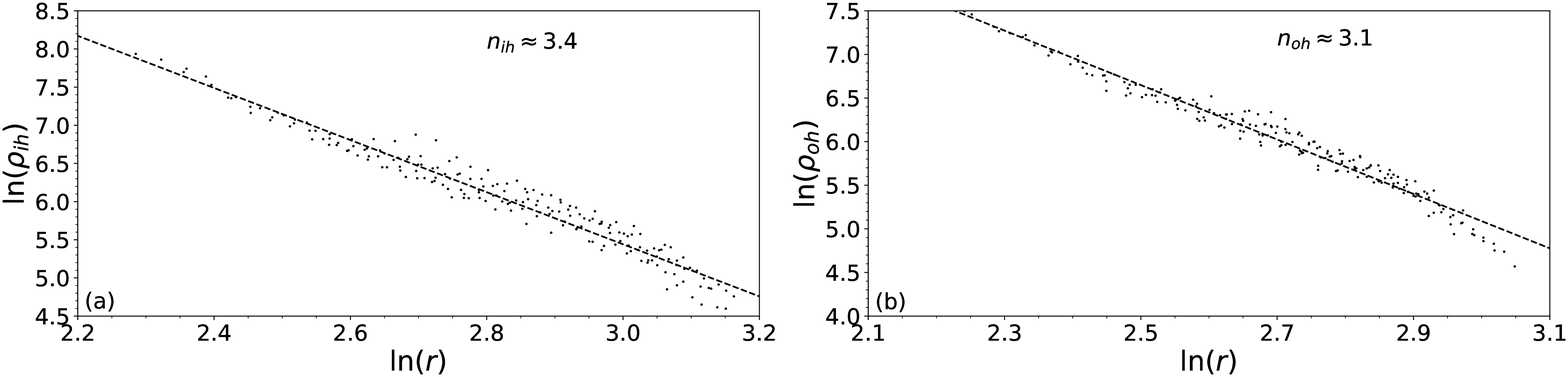}}
\end{minipage}
\end{center}
\caption{$\ln(\rho)$ vs. $\ln(r)$. Panel (a): The case of inner halo. Panel (b): The case of outer halo. In each panel, the asterisks are the data points, each of which corresponds to the a discretized cell in the considered $R$-$Z$ region (Eq. \ref{eq_RZ_region}). The dash line is obtained from linearly fitting these data points, with the slope identified to be the minus sign of power-law index $n$ according to Eq. (\ref{eq_halo_model_linear}).}
\label{fig_n}
\end{figure*}

\par In Figure \ref{fig_q}, we plot the $var_{sum}$ versus $q$ for two halo component. The best-fit values of $q$ for both inner halo and outer halo are
\begin{align}
q_{ih}\approx 0.49 \hspace{0.5cm}\text{and}\hspace{0.5cm} q_{oh}\approx 0.61 \text{,}
\end{align}
which are consistent with the typical range $[0.5,\,1.0]$ in previous works \citep{Siegel2002, Du2003, Du2006, Bilir2008, Yaz2010, Chang2011, Jia2014}.
It also indicates that the outer halo is more spherical than inner halo.

\par After getting the best-fit value of $q$ for both halos, we turn to the estimation of power-law index $n$. As shown in Figure \ref{fig_n}, the $\ln(\rho_D)$ is plotted as a function of $\ln(r)$ for each halo. In each panel of Figure \ref{fig_n}, the asterisks are the data points corresponding to the discretized cells in considered $R$-$Z$ region, and the data behavior can be captured roughly by a the dash line which is obtained from linear regression with the data points. According to linear model of halos (Eq. \ref{eq_halo_model_linear}), 
the power-law index $n$ for the halo can be derived, 
\begin{align}
n_{ih}\approx 3.4 \hspace{0.5cm}\text{and}\hspace{0.5cm} n_{oh}\approx 3.1 \text{.}
\end{align}
This manifests that the change of stellar number density with distance in the inner halo is more steep than outer halo.  Let's now go back to Figure \ref{fig_CMD}, it shows that only 69\% of F/G main-sequence stars are selected in the region with $7<R<14\,{\rm kpc}$ and $3<Z<10\,{\rm kpc}$, in order to ensure the accuracy of $u$-band magnitude. We now discuss how this  influence the best-fit value of power-law index. The incompleteness of sample stars becomes more severe as the radius $r$ gets larger, which can lead to the best-value of power-law index larger than the expected value. As shown in Figure \ref{fig_n}, we can also see that at the large end of $\ln(r)$, the data points are under dash line derived from linear regression, which can be due to the incompleteness of sample stars. So, if the incompleteness of sample stars was corrected, power index $n$ could be expected a little smaller.  Additionally, if we perform the correction to classify a small fraction of metal-rich stars to halo, as indicated by the constant $C_D$ in Eq. (\ref{eq_disk_model}), $n$ would be further smaller, and is likely lying in $[2.5,\,3.0]$ which is the typical range for power-law index from previous works.

\begin{figure}
\begin{center}
\begin{minipage}[t]{1.00\hsize}
\resizebox{1.0\hsize}{!}{\includegraphics{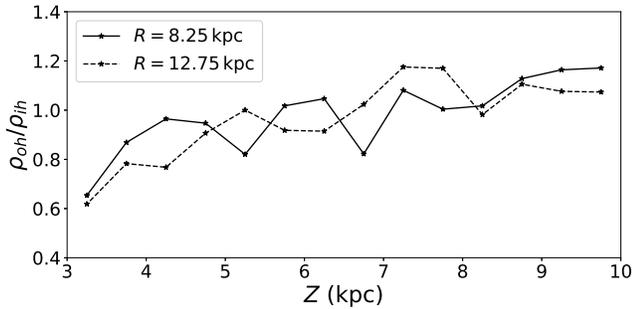}}
\end{minipage}
\end{center}
\caption{The ratio of stellar number density of outer halo over that of inner halo which varies as a function of $Z$ with $R$ fixed. }
\label{fig_rho_ratio}
\end{figure}

\par Now that the halo is comprised of two distinct but overlapping components. It is interesting to study the fraction of each component which is reasonably varies with position. As shown in Figure \ref{fig_rho_ratio}, we compute the stellar number density ratio between outer halo and inner halo, and see how it varies in vertical direction. Two cases with $R=8.25\,{\rm kpc}$ and $R=12.75\,{\rm kpc}$ are given.  Despite some irregularities, we can clearly see the trend that the fraction of outer halo stars become larger with the vertical distance.

\section{Conclusions and Perspectives}

\par Based on the Monte-Carlo-based method for estimating photometric MDFs, we fit the MDFs with three Gaussians model, with peaks at ${\rm [Fe/H]}$=$-0.68$, $-1.38$, and $-1.90$, associated with the thick disk, inner halo, and outer halo, respectively.  
This enables a successful separation of  three main components in a statistical way. The vertical metallicity gradient within $1<Z<5\,{\rm kpc}$ is ${\rm d}\langle{\rm [Fe/H]}\rangle/{\rm d}Z\approx -0.19\,{\rm dex}\cdot{\rm kpc}^{-1}$ around $R=8.25\,{\rm kpc}$. But the mean radial gradient is almost negligible. 
The density profile of thick disk can be modeled with a modified double exponential model decaying to a constant at far distance, which are plausibly explained to belong to Galactic halo.   The scale height and length are well determined, $H\approx 1.13\,{\rm kpc}$ and $L\approx 3.63\,{\rm kpc}$. The halos are described with power-law ellipsoid  and the axis ratios of both inner halo and outer halo are $q_{ih}\approx 0.49$ and $q_{oh}\approx 0.61$, respectively. The halo power-law indices are $n_{ih}\approx 3.4$ and $n_{oh}\approx 3.1$. It shows that the outer halo is more spherical and less steeper than inner halo. These results are in consistent with the results determined from star-counts method in previous studies.  

\par This work sets up a connection between studies of metallicity distribution and spatial structure of the Galaxy. An explicit correlation between the two aspects has been shown, and further this correlation can be used to constrain Galactic model that offers essential clues to the galactic formation and evolution. Following that, a more comprehensive study is hoped to be carried out in the future.

\section*{Acknowledgements}
\par We thank the anonymous referee for his/her comments that greatly improve this paper. This work is financially supported by the China Scholarship Council under the Grant No. 201606950037.

\par Funding for SDSS-III has been provided by the Alfred P. Sloan Foundation, the Participating Institutions, the National Science Foundation, and the U.S. Department of Energy Office of Science. The SDSS-III web site is \emph{http://www.sdss3.org/}. SDSS-III is managed by the Astrophysical Research Consortium for the Participating Institutions of the SDSS-III Collaboration including the University of Arizona, the Brazilian Participation Group, Brookhaven National Laboratory, Carnegie Mellon University, University of Florida, the French Participation Group, the German Participation Group, Harvard University, the Instituto de Astrofisica de Canarias, the Michigan State/Notre Dame/JINA Participation Group, Johns Hopkins University, Lawrence Berkeley National Laboratory, Max Planck Institute for Astrophysics, Max Planck Institute for Extraterrestrial Physics, New Mexico State University, New York University, Ohio State University, Pennsylvania State University, University of Portsmouth, Princeton University, the Spanish Participation Group, University of Tokyo, University of Utah, Vanderbilt University, University of Virginia, University of Washington, and Yale University.

\par The SCUSS is funded by the Main Direction Program of Knowledge Innovation of Chinese Academy of Sciences (No. KJCX2-EW-T06). It is also an international cooperative project between the National Astronomical Observatories, Chinese Academy of Sciences and Steward Observatory, University of Arizona, USA. Technical support and observational assistances of the Bok telescope are provided by Steward Observatory. The project is managed by the National Astronomical Observatory of China and Shanghai Astronomical Observatory.

\end{document}